\documentclass{article}

\usepackage{amsmath,amssymb}
\usepackage{graphicx}
\usepackage{color}
\usepackage[all]{xy}

\newtheorem{theorem}{Theorem}
\newtheorem{example}{Example}
\newtheorem{lemma}{Lemma}

\newtheorem{definition}{Definition}
\newtheorem{proposition}{Proposition}
\newtheorem{remark}{Remark}

\newcommand{\done}{\hfill $\Box$ }

\newcommand{\ls}[1]
    {\dimen0=\fontdimen6\the\font\lineskip=#1\dimen0
     \advance\lineskip.5\fontdimen5\the\font
     \advance\lineskip-\dimen0
     \lineskiplimit=0.9\lineskip
     \baselineskip=\lineskip
     \advance\baselineskip\dimen0
     \normallineskip\lineskip\normallineskiplimit\lineskiplimit
     \normalbaselineskip\baselineskip
     \ignorespaces}

\begin{document}

\bibliographystyle{abbrv}

\title{On the Dual of the Coulter-Matthews Bent Functions}

\author{Honggang Hu, Qingsheng Zhang, and Shuai Shao\\
School of Information Science and Technology\\
University of Science and Technology of China\\
Hefei, China, 230027\\
Email. hghu2005@ustc.edu.cn }

\date{}
 \maketitle

\thispagestyle{plain}
\setcounter{page}{1}

\begin{abstract}
For any bent function, it is very interesting to determine its dual function because the dual function is also bent in certain cases.
For $k$ odd and $\gcd(n, k)=1$, it is known that the Coulter-Matthews bent function $f(x)=Tr(ax^{\frac{3^k+1}{2}})$ is weakly regular
bent over $\mathbb{F}_{3^n}$, where $a\in\mathbb{F}_{3^n}^{*}$, and $Tr(\cdot):\mathbb{F}_{3^n}\rightarrow\mathbb{F}_3$ is the trace function.
In this paper, we investigate the dual function of $f(x)$, and dig out an universal formula. In particular, for two cases,
we determine the formula explicitly: for the case of $n=3t+1$ and $k=2t+1$ with $t\geq 2$, the dual function is given by
$$Tr\left(-\frac{x^{3^{2t+1}+3^{t+1}+2}}{a^{3^{2t+1}+3^{t+1}+1}}-\frac{x^{3^{2t}+1}}{a^{-3^{2t}+3^{t}+1}}+\frac{x^{2}}{a^{-3^{2t+1}+3^{t+1}+1}}\right);$$
and for the case of $n=3t+2$ and $k=2t+1$ with $t\geq 2$, the dual function is given by
$$Tr\left(-\frac{x^{3^{2t+2}+1}}{a^{3^{2t+2}-3^{t+1}+3}}-\frac{x^{2\cdot3^{2t+1}+3^{t+1}+1}}{a^{3^{2t+2}+3^{t+1}+1}}+\frac{x^2}{a^{-3^{2t+2}+3^{t+1}+3}}\right).$$
As a byproduct, we find two new classes of ternary bent functions with only three terms. Moreover,
we also prove that in certain cases $f(x)$ is regular bent.
\end{abstract}


\ls{1.5}
%===========================================================================
%===========================================================================
\section{Introduction}\label{sec_intro}

Boolean bent functions have the maximum Hamming distance to
the set of all affine functions, which was first introduced by
Rothaus in 1976 \cite{Rothaus76}. Because of many applications in cryptography,
coding theory and communications \cite{GG05,HK98}, bent functions have been an active research issue for a long time.
In 1985, Kumar, Scholtz and Welch generalized the concept of Boolean bent functions to the case of functions
over integer residue rings \cite{KuScWe85}. Compared with Boolean bent functions, generalized bent functions
are much more complicated. However, there are some known methods to construct bent functions over finite fields \cite{HeKh06_1,HeKh07,HeHoKhWaXi09,HeKh10_4,Ho08}.

Let $\mathbb{F}_{p^n}$ be the finite field of $p^n$ elements, where $p$ is a prime number and $n\geq 1$.
For a function $f(x)$ from $\mathbb{F}_{p^n}$ to $\mathbb{F}_p$, its Walsh transform $\widehat{f}(\lambda)$ is defined by
$$\widehat{f}(\lambda)=\sum_{x\in \mathbb{F}_{p^n}}\omega_p^{f(x)-Tr(\lambda x)},$$
where $Tr(\cdot)$ is the trace function from $\mathbb{F}_{p^n}$ to $\mathbb{F}_p$, and $\omega_p=e^{2\pi i/p}$ is the complex primitive $p$-th root of unity.
The inverse Walsh transform of $f(x)$ is defined by
$$\omega_p^{f(x)}=\frac{1}{p^n}\sum_{\lambda\in \mathbb{F}_{p^n}}\widehat{f}(\lambda)\omega_p^{Tr(\lambda x)}.$$
If the following equality $$|\widehat{f}(\lambda)|^2=p^n$$
holds for all $\lambda\in \mathbb{F}_{p^n}$, then $f(x)$ is called a $p$-ary bent function (or a generalized bent function over $\mathbb{F}_{p^n}$).

A $p$-ary bent function $f(x)$ is called regular if there exists a $p$-ary function $g(x)$ from $\mathbb{F}_{p^n}$ to $\mathbb{F}_p$ such that
$$\widehat{f}(\lambda)=p^{\frac{n}{2}}\omega_p^{g(\lambda)}$$
holds for all $\lambda\in \mathbb{F}_{p^n}$. A $p$-ary bent function $f(x)$ is called weakly regular if there exists a complex number $u$
with $|u|=1$ such that $$\widehat{f}(\lambda)=up^{\frac{n}{2}}\omega_p^{g(\lambda)}$$
holds for all $\lambda\in \mathbb{F}_{p^n}$, where $g(x)$ is a $p$-ary function from $\mathbb{F}_{p^n}$ to $\mathbb{F}_p$. In this case,
$g(x)$ is said to be the dual function of $f(x)$, which is also weakly regular bent. For the constant $u$, it is known that it can be equal
to only four values: $\pm1, \pm i$ \cite{HeKh06_1}.

The case of ternary bent function is particularly interesting. Let $d=\frac{3^k+1}{2}$, where $k$ is odd, and $\gcd(n, k)=1$.
In 1997, Coulter and Matthews showed that $x^d$ is a planar function over $\mathbb{F}_{3^n}$ \cite{CM97}. As a consequence,
$Tr(ax^d)$ is bent over $\mathbb{F}_{3^n}$ for any $a\in \mathbb{F}_{3^n}^{*}$, which is called the Coulter-Matthews bent function \cite{HeHoKhWaXi09}.
In 2006, Helleseth and Kholosha conjectured that the Coulter-Matthews bent functions are weakly regular bent \cite{HeKh06_1,HeHoKhWaXi09}. In 2008, this
conjecture was proved in two special cases \cite{Ho08}. Later, this conjecture was completely settled in 2009 \cite{HeHoKhWaXi09}. In 2006,
Helleseth and Kholosha also conjectured another kind of monomial bent functions over $\mathbb{F}_{3^n}$ \cite{HeKh06_1}. The first part of this conjecture
was proved in 2009 \cite{HeHoKhWaXi09}, and the second part of this conjecture was proved in 2012 \cite{GHHK12}. In addition to the complete proof,
in \cite{GHHK12}, a specific formula for the dual function was also presented via certain deep tools in cyclotomic number theory.

For any bent function, it is pretty interesting to determine its dual function because the dual function is also bent. In this paper, we investigate the
dual function of the Coulter-Matthews bent functions. Via Stickelberger's theorem and certain tricks, we find an universal formula for the dual function.
Using new combinatorial methods we develop in this paper, for two cases, we determine the formula explicitly: if $n=3t+1$ and $k=2t+1$ with $t\geq 2$,
then the dual function is given by
$$g(x)=Tr\left(-\frac{x^{3^{2t+1}+3^{t+1}+2}}{a^{3^{2t+1}+3^{t+1}+1}}-\frac{x^{3^{2t}+1}}{a^{-3^{2t}+3^{t}+1}}+\frac{x^{2}}{a^{-3^{2t+1}+3^{t+1}+1}}\right);$$
if $n=3t+2$ and $k=2t+1$ with $t\geq 2$, then the dual function is given by
$$g(x)=Tr\left(-\frac{x^{3^{2t+2}+1}}{a^{3^{2t+2}-3^{t+1}+3}}-\frac{x^{2\cdot3^{2t+1}+3^{t+1}+1}}{a^{3^{2t+2}+3^{t+1}+1}}+\frac{x^2}{a^{-3^{2t+2}+3^{t+1}+3}}\right).$$
From the viewpoint of bent functions, we dig out two classes of ternary bent functions with only three terms,
which have never been reported in the literature. From the viewpoint of character sums, we determine the values of two kinds
of character sums over $\mathbb{F}_{3^n}$ exactly, which is very interesting itself.

This paper is organized as follows. In Section \ref{sec_pre}, we give some necessary notation and background. In Section \ref{sec_formula}, we
present the universal formula for the dual function. Sections \ref{sec_case1} and \ref{sec_case2} consider two specific cases in details, respectively.
Finally, Section \ref{sec_con} concludes this paper.

%===========================================================================
%===========================================================================
\section{Preliminaries}\label{sec_pre}

%===========================================================================
\subsection{Cyclotomic Cosets Modulo $3^n-1$}

A cyclotomic coset $C_s$ modulo $3^n-1$ is defined by
$$C_s=\{s, 3s, ..., 3^{n_s-1}s\},$$
where $n_s$ is the smallest positive integer such that $s\equiv
s3^{n_s}(\mbox{mod }3^n-1)$. Note that $n_s|n$. The subscript $s$ is selected as the
smallest integer in $C_s$, and $s$ is said to be the coset leader of $C_s$.  For example, for $n=3$, the cyclotomic cosets modulo 26 are:
$$C_0=\{0\}, C_1=\{1, 3, 9\}, C_2=\{2, 6, 18\}, C_4=\{4, 10, 12\}, C_5=\{5, 15, 19\}, C_7=\{7, 11, 21\}$$
$$C_8=\{8, 20, 24\}, C_{13}=\{13\}, C_{14}=\{14, 16, 22\}, C_{17}=\{17, 23, 25\}$$
where $\{0, 1, 2, 4, 5, 7, 8, 13, 14, 17\}$ are coset leaders modulo 26.

\begin{proposition}[Trace Representation \cite{GG05}]\label{prop_trace}
Any nonzero function $f(x)$ from $\mathbb{F}_{3^n}$ to
$\mathbb{F}_3$ can be represented as
$$f(x)=\sum_{k\in \Gamma(n)}Tr_1^{n_k}(F_kx^k)+F_{3^n-1}x^{3^n-1}, F_k\in \mathbb{F}_{3^{n_k}}, F_{3^n-1}\in \mathbb{F}_3$$
where $\Gamma(n)$ is the set consisting of all coset leaders modulo
$3^n-1$, $n_k|n$ is the size of the coset $C_k$, and $Tr_1^{n_k}(x)$
is the trace function from $\mathbb{F}_{3^{n_k}}$ to $\mathbb{F}_3$.
\end{proposition}

%===========================================================================
\subsection{Gauss Sums and Stickelberger's Theorem}

Let $\mathbb{F}_q$ be the finite field of $q$ elements, where $q=p^n$, and $p$ is a prime number. Let $\psi$
be the mapping defined by
$$\psi(x)=\omega_p^{Tr(x)}.$$
Then $\psi$ is an additive character of $\mathbb{F}_q$. Let $\chi$ be a multiplicative character of $\mathbb{F}_q^{*}$.
For the convenience, we can extend $\chi$ to $\mathbb{F}_q$ by defining $\chi(0)=0$. From now on, for simplicity,
the multiplicative character set of $\mathbb{F}_q^{*}$ is denoted by $\widehat{\mathbb{F}_q^{*}}$ .

\begin{definition}[\cite{LN83}]
For any multiplicative character $\chi$ over $\mathbb{F}_q$, the
Gauss sum $G(\chi)$ over $\mathbb{F}_q$ is defined by
$$G(\chi)=\sum_{x\in F_q}\psi(x)\chi(x).$$
\end{definition}

\begin{lemma}[\cite{LN83}]\label{lem_gauss}
For any multiplicative character $\chi$ over $\mathbb{F}_q$, we have
$$G(\overline{\chi})=\chi(-1)\overline{G(\chi)}\mbox{ and }G(\chi^p)=G(\chi).$$
If $\chi$ is trivial, then $G(\chi)=-1$. Moreover, if $\chi$ is
nontrivial, then
$$G(\chi)\overline{G(\chi)}=q.$$
\end{lemma}

\begin{figure}
\centering
$$\xymatrix{
& \mathbb{Z}[\omega_p, \omega_{q-1}] & \\
  \mathbb{Z}[\omega_p] \ar[ur]^{(\pi)=\mathcal{Q}_1\mathcal{Q}_2\cdots  \mathcal{Q}_t}  &  &    \mathbb{Z}[\omega_{q-1}] \ar[ul]_{\mathfrak{p}_i=\mathcal{Q}_i^{p-1}}    \\
                & \ar[ul]^{(p)=(\pi)^{p-1}} \mathbb{Z}    \ar[ur]_{(p)=\mathfrak{p}_1\mathfrak{p}_2\cdots  \mathfrak{p}_t}
                }$$
\caption{Prime Ideal Factorization}\label{fig_primeideal}
\end{figure}
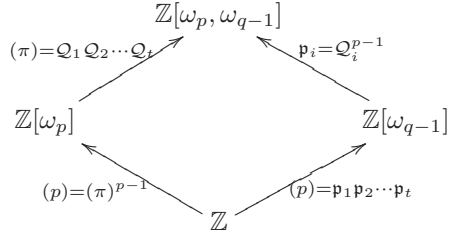

In algebraic integer rings, the factorization of prime ideals is interesting and very useful.
Firstly, $(p)$ is a prime ideal in $\mathbb{Z}$. Let $\pi=\omega_p-1$. Then $(\pi)$ is a prime ideal in
$\mathbb{Z}[\omega_p]$, and $(p)=(\pi)^{p-1}$. Furthermore, $(\pi)$ can be factored into the product of different prime ideals
in $\mathbb{Z}[\omega_p, \omega_{q-1}]$, i.e.,
$(\pi)=\mathcal{Q}_1\mathcal{Q}_2\cdots  \mathcal{Q}_t$, where $\mathcal{Q}_i$ are
prime ideals in $\mathbb{Z}[\omega_p, \omega_{q-1}]$, and $t=\phi(p^n-1)/n$. Therefore, $(p)=(\mathcal{Q}_1\mathcal{Q}_2\cdots  \mathcal{Q}_t)^{p-1}$ in
$\mathbb{Z}[\omega_p, \omega_{q-1}]$. On the other hand, $(p)$ can be factored into the product of $t$ different prime ideals in
$\mathbb{Z}[\omega_{q-1}]$, i.e., $(p)=\mathfrak{p}_1\mathfrak{p}_2\cdots  \mathfrak{p}_t$, where $\mathfrak{p}_i$ are
prime ideals in $\mathbb{Z}[\omega_{q-1}]$. For each $\mathfrak{p}_i$, it can be written in the form of $(p-1)$-th power of
a prime ideal in $\mathbb{Z}[\omega_p,\omega_{q-1}]$. Without loss of generality, let $\mathfrak{p}_i=\mathcal{Q}_i^{p-1}$.
The reader is referred to Figure \ref{fig_primeideal} for the relation among $(p),\mathfrak{p}_i$, and $\mathcal{Q}_i$.

For each $\mathcal{Q}_i$, because $[\mathbb{Z}[\omega_p, \omega_{q-1}]/\mathcal{Q}_i:\mathbb{Z}/(p)]=n$, we have
$$\mathbb{Z}[\omega_p,\omega_{q-1}]/\mathcal{Q}_i\cong \mathbb{F}_q.$$
From now on, we fix one prime ideal
$\mathcal{Q}_i$ which is denoted by $\mathcal{Q}$ for simplicity. There is one special multiplicative character $\chi$
on $\mathbb{F}_q$ called the Teichm\"{u}ller character, which satisfies
$$\chi(x)(\mbox{mod }\mathcal{Q})=x.$$
For simplicity, from now on we denote the Teichm\"{u}ller character by $\chi_\mathfrak{p}$.

For any $0\leq k<q-1$, let $k=\sum_{i=0}^{n-1}k_ip^{i}$ be the
$p$-adic representation of $k$, where $0\leq k_i<p$ for
$i=0,1,\dots,n-1$. Let $\mathrm{wt}(k)=\sum_{i=0}^{n-1}k_i$, and $\sigma(k)=\prod_{i=0}^{n-1}k_i!$.
Furthermore, for any $j$, we use $\mathrm{wt}(j)$ and $\sigma(j)$ to denote
$\mathrm{wt}(\overline{j})$ and $\sigma(\overline{j})$ respectively, where $0\leq \overline{j}<q-1$
and $j\equiv \overline{j}\ (\bmod\;q-1)$.

\begin{theorem}[Stickelberger's Theorem \cite{Lang78}]\label{thm_Stickelberger}
For any $0<k<q-1$, we have
$$G(\chi_\mathfrak{p}^{-k})\equiv -\frac{\pi^{wt(k)}}{\sigma(k)}(\mbox{mod }\pi^{wt(k)+p-1}).$$
\end{theorem}

The following lemma is well known and extremely powerful.

\begin{lemma}\label{lem_gauss_trace}
For any $y\in \mathbb{F}_{q}^{*}$, we have
$$\omega_p^{Tr(y)}=\frac{1}{q-1}\sum_{\chi\in\widehat{\mathbb{F}_{q}^{*}}}G(\chi)\overline{\chi}(y).$$
\end{lemma}

Stickelberger's theorem has played a significant role in the proof of some important conjectures:
Welch, Niho, and Lin conjectures \cite{Mc72,CaChDo00,HoXi01,HSGH14}.

%===========================================================================
\subsection{The Ternary Modular Add-With-Carry Algorithm}

For any three integers $0\leq x, y, z<3^n-1$, let $x=\sum_{i=0}^{n-1}x_i3^{i}$, $y=\sum_{i=0}^{n-1}y_i3^{i}$,
and $z=\sum_{i=0}^{n-1}z_i3^{i}$ be the ternary representations of $x, y, z$ respectively, where $0\leq x_i, y_i, z_i<3$ for
$i=0,1,\dots,n-1$. Suppose that $z\equiv x+y\mbox{ mod }3^n-1$. Then there exists a unique integer sequence
$\overrightarrow{c}=c_0, c_1, ..., c_{n-1}$ with $c_i=0$ or $1$ for $i=0, 1, ..., n-1$ such that
$$z_i+3c_i=x_i+y_i+c_{i-1},\ 0\leq i\leq n-1,$$
where the subscript takes the value modulo $n$. From now on, all subscripts take the value modulo $n$.

Put $\mathrm{wt}(\overrightarrow{c})=c_0+c_1+...+c_{n-1}$. Then we have
$$\mathrm{wt}(x)+\mathrm{wt}(y)=\mathrm{wt}(z)+2\mathrm{wt}(\overrightarrow{c})\geq\mathrm{wt}(x+y).$$

\begin{lemma}\label{label_xy=}
With notations as above, $\mathrm{wt}(x)+\mathrm{wt}(y)=\mathrm{wt}(x+y)$ if and only if $x_i+y_i\leq 2$ for $0\leq i\leq n-1$,
and there exists $j$ with $0\leq j\leq n-1$ such that $x_j+y_j<2$.
In particular, $2\mathrm{wt}(x)=\mathrm{wt}(2x)$ if and only if $x_i\neq 2$ for $0\leq i\leq n-1$,
and $x\neq \frac{3^n-1}{2}$.
\end{lemma}
{\bf Proof. }Because $\mathrm{wt}(x)+\mathrm{wt}(y)=\mathrm{wt}(z)+2\mathrm{wt}(\overrightarrow{c})$, we have
$\mathrm{wt}(x)+\mathrm{wt}(y)=\mathrm{wt}(x+y)$ if and only if $\mathrm{wt}(\overrightarrow{c})=0$ and $x+y\neq 3^n-1$. The result follows.
\done

\begin{lemma}\label{lem_xy+2}
With notations as above, $\mathrm{wt}(x)+\mathrm{wt}(y)=\mathrm{wt}(x+y)+2$ if and only if there exists $j$
with $0\leq j\leq n-1$ such that $x_j+y_j\geq 3$ and $x_{j+1}+y_{j+1}\leq 1$, and for $0\leq i\leq n-1$ with $i\neq j, j+1$, $x_i+y_i\leq 2$.
In particular, $2\mathrm{wt}(x)=\mathrm{wt}(2x)+2$ if and only if there exists $j$
with $0\leq j\leq n-1$ such that $x_j=2$ and $x_{j+1}=0$, and for $0\leq i\leq n-1$ with $i\neq j, j+1$, $x_i\leq 1$.
\end{lemma}
{\bf Proof. }Because $\mathrm{wt}(x)+\mathrm{wt}(y)=\mathrm{wt}(z)+2\mathrm{wt}(\overrightarrow{c})$, we have
$\mathrm{wt}(x)+\mathrm{wt}(y)=\mathrm{wt}(x+y)+2$ if and only if $\mathrm{wt}(\overrightarrow{c})=1$. The result follows.
\done

%===========================================================================
%===========================================================================
\section{An Universal Formula}\label{sec_formula}

Henceforth, let $q=3^n$, where $n$ is a positive integer.
Let $\pi=\omega_3-1$. Then $\pi=-\sqrt{3}i\omega_3^2$.

Now we study $\sum_{x\in \mathbb{F}_q}\omega_3^{Tr(ax^d+\lambda x)}$ for any $a\in \mathbb{F}_q^*$, $\lambda\in \mathbb{F}_q$,
where $d=(3^k+1)/2$ with $k$ odd and $\gcd(n, k)=1$. It is known that $|\sum_{x\in \mathbb{F}_q}\omega_3^{Tr(ax^d+\lambda x)}|=3^{n/2}$ \cite{FL07,HeHoKhWaXi09}.

For any $j$, let $\overline{j}$ be the residue of $j$ modulo $3^n-1$, i.e., $j\equiv \overline{j}\ (\bmod\;3^n-1)$ and $0\leq \overline{j}<3^n-1$.
Let $\overline{j}=\sum_{i=0}^{n-1}j_i3^{i}$ be the ternary representation of $\overline{j}$, where $j_i\in\{0, 1, 2\}$ for $0\leq i<n$.
For simplicity, we write $\overline{j}$ in the form of $j_{n-1}j_{n-2}\cdots j_1j_0$, i.e., $j_{n-1}j_{n-2}\cdots j_1j_0$ is the ternary representation of $\overline{j}$. If $\overline{j}\neq 0$, then $\overline{-j}=\sum_{i=0}^{n-1}\widehat{j}_i3^{i}$, where $\widehat{j}_i=2-j_i$ for $0\leq i<n$.

\begin{lemma}\label{lem_-j}
With notations as above, if $\overline{j}\neq \frac{3^n-1}{2}$, then $2\mathrm{wt}(-j)=\mathrm{wt}(-2j)$ if and only if $j_i\neq0$ for $0\leq i<n$.
\end{lemma}
{\bf Proof. }By Lemma \ref{label_xy=}, $2\mathrm{wt}(-j)=\mathrm{wt}(-2j)$ if and only if $\widehat{j}_i\neq 2$ for $0\leq i<n$
which means that $j_i\neq0$ for $0\leq i<n$. \done

\begin{lemma}\label{lem_hi_1}
With notations as above, for any $0<j<3^n-1$ with $j\neq(3^n-1)/2$ satisfying $\mathrm{wt}(j)+\mathrm{wt}(3^kj)=\mathrm{wt}((3^k+1)j)$, let $h=h_{n-1}h_{n-2}\cdots h_1h_0=\overline{(3^k+1)j}$. Let $j_{n-1}j_{n-2}\cdots j_1j_0$ be the ternary representation of $j$. If there exists
$0\leq i<n$ such that $j_i=1$, then there exist $0\leq a<b<n$ such that $h_a=h_b=1$.
\end{lemma}
{\bf Proof. }Without loss of generality, we may assume that $j_0=1$. Suppose that $h_i\neq1$ for any $0\leq i<n$.
Because $\mathrm{wt}(j)+\mathrm{wt}(3^kj)=\mathrm{wt}((3^k+1)j)$, we have $j_{ik}=1$ for any $i\geq0$. Thus,
$$j_{n-1}j_{n-2}\cdots j_1j_0=\underbrace{11\cdots 11}_n,$$
which means that $j=(3^n-1)/2$. This is a contradiction. Hence, there exists $0\leq a<n$ such that $h_a=1$.
Because $h$ is even, there exists $0\leq b<n$ with $a\neq b$ such that $h_b=1$. Without loss of generality, we may assume $a<b$.\done

The following two lemmas can be checked easily.

\begin{lemma}\label{lem_000}
For any $m\geq 0$, we have
$$\frac{1\overbrace{00\cdots00}^{m}1}{2}=\overbrace{11\cdots11}^{m}2.$$
\end{lemma}

\begin{lemma}\label{lem_222}
For any $m\geq 0$, we have
$$\frac{1\overbrace{22\cdots22}^{m}1}{2}=\overbrace{22\cdots22}^{m+1}.$$
\end{lemma}

\begin{lemma}\label{lem_n}
With notations as above, $\mathrm{wt}(j)+\mathrm{wt}(-jd)\geq n$ for any $0<j<3^n-1$, and $j=(3^n-1)/2$ is the only value such that $\mathrm{wt}(j)+\mathrm{wt}(-jd)=n$.
Moreover, $\mathrm{wt}(j)+\mathrm{wt}(-jd)=n+1$ if and only if one of two following conditions holds:

1) $\mathrm{wt}(j)+\mathrm{wt}(3^kj)=\mathrm{wt}((3^k+1)j)$ and $2\mathrm{wt}(-jd)=\mathrm{wt}(-(3^k+1)j)+2$;

2) $\mathrm{wt}(j)+\mathrm{wt}(3^kj)=\mathrm{wt}((3^k+1)j)+2$ and $2\mathrm{wt}(-jd)=\mathrm{wt}(-(3^k+1)j)$.
\end{lemma}
{\bf Proof. }
If $j=(3^n-1)/2$, then we have
$$\mathrm{wt}(j)+\mathrm{wt}(-jd)=\mathrm{wt}(j)+\mathrm{wt}(0)=\mathrm{wt}(j)=n.$$
For any $0<j<3^n-1$ with $j\neq(3^n-1)/2$, it holds that $\mathrm{wt}(-jd)\neq 0$ because $\gcd(d, 3^n-1)=2$.
Thus, in this case, we have
\begin{eqnarray*}
\mathrm{wt}(j)+\mathrm{wt}(-jd)&=&\frac{1}{2}(\mathrm{wt}(j)+\mathrm{wt}(j)+2\mathrm{wt}(-jd))\\
&=&\frac{1}{2}(\mathrm{wt}(j)+\mathrm{wt}(3^kj)+2\mathrm{wt}(-jd))\\
&\geq&\frac{1}{2}(\mathrm{wt}((3^k+1)j)+\mathrm{wt}(-(3^k+1)j))\\
&=&n.
\end{eqnarray*}
The equality holds if and only if $$\mathrm{wt}(j)+\mathrm{wt}(3^kj)=\mathrm{wt}((3^k+1)j)\mbox{ and }2\mathrm{wt}(-jd)=\mathrm{wt}(-(3^k+1)j).$$
It is straightforward to see that  $\mathrm{wt}(j)+\mathrm{wt}(-jd)=n+1$ if and only if one of two conditions below holds:

1) $\mathrm{wt}(j)+\mathrm{wt}(3^kj)=\mathrm{wt}((3^k+1)j)$ and $2\mathrm{wt}(-jd)=\mathrm{wt}(-(3^k+1)j)+2$;

2) $\mathrm{wt}(j)+\mathrm{wt}(3^kj)=\mathrm{wt}((3^k+1)j)+2$ and $2\mathrm{wt}(-jd)=\mathrm{wt}(-(3^k+1)j)$.

For any $0<j<3^n-1$ with $j\neq(3^n-1)/2$, which satisfies $\mathrm{wt}(j)+\mathrm{wt}(3^kj)=\mathrm{wt}((3^k+1)j)$,
our goal is to prove that $2\mathrm{wt}(-jd)>\mathrm{wt}(-(3^k+1)j)$. Let $h=h_{n-1}h_{n-2}\cdots h_1h_0=\overline{(3^k+1)j}$.
Then $h\neq 0$. Let $u=u_{n-1}u_{n-2}\cdots u_1u_0=\overline{jd}$. Then $u=\frac{h}{2}$, or $u=(\frac{h}{2}+\frac{3^n-1}{2})(\bmod\;3^n-1)$.

i) $u=\frac{h}{2}$

Suppose that $h_i\neq1$ for $0\leq i<n$. Then there exists $0\leq a<n$ such that $h_a=0$ because $h\neq0$.
It follows that $u_a=0$, which means that $2\mathrm{wt}(-u)>\mathrm{wt}(-2u)$ by Lemma \ref{lem_-j}.

Suppose that there exists $0\leq a<n$ such that $h_a=1$. There there exists $0\leq b<n$ with $a\neq b$ such that $h_b=1$ because $h$ is even.
Without loss of generality, let $b>a$. By Lemmas \ref{lem_000} and \ref{lem_222}, we have $u_b=0$,
which means that $2\mathrm{wt}(-u)>\mathrm{wt}(-2u)$ by Lemma \ref{lem_-j}.

ii) $u=(\frac{h}{2}+\frac{3^n-1}{2})(\bmod\;3^n-1)$

Let $j_{n-1}j_{n-2}\cdots j_1j_0$ be the ternary representation of $j$. If $j_i\neq1$ for any $0\leq i<n$, let $\widehat{j}=j/2$.
Suppose that $\widehat{j}_{n-1}\widehat{j}_{n-2}\cdots \widehat{j}_1\widehat{j}_0$ is the ternary representation of $\widehat{j}$.
Then $\widehat{j}_i=j_i/2$. Because $\mathrm{wt}(j)+\mathrm{wt}(3^kj)=\mathrm{wt}((3^k+1)j)$, we have $\mathrm{wt}(\widehat{j})+\mathrm{wt}(3^k\widehat{j})=\mathrm{wt}((3^k+1)\widehat{j})$. It follows that $u=\overline{jd}=\overline{(3^k+1)\widehat{j}}<(3^n-1)/2$. Therefore, $h=\overline{(3^k+1)j}=2u$. Hence, this case is impossible, and
there exists $0\leq i<n$ such that $j_i=1$. By Lemma \ref{lem_hi_1}, there exist $0\leq a<b<n$ such that $h_a=h_b=1$. By Lemmas \ref{lem_000} and \ref{lem_222}, we have $u_a=0$, which means that $2\mathrm{wt}(-u)>\mathrm{wt}(-2u)$ by Lemma \ref{lem_-j}. \done

\begin{lemma}[\cite{LN83}]
Let $\eta$ be the quadratic character of $\mathbb{F}_q$. Then $G(\eta)=(-1)^{n+1}i^n3^{n/2}$.
\end{lemma}

\begin{remark}
Note that $G(\eta)=-\omega_3^n\pi^n$.
\end{remark}

\begin{theorem}\label{thm_lambda_0}
For any $a\in \mathbb{F}_q^*$, $\sum_{x\in \mathbb{F}_q}\omega_3^{Tr(ax^d)}=(-1)^{n+1}\eta(a)i^n3^{n/2}$.
\end{theorem}
{\bf Proof. }Because $\gcd(d, 3^n-1)=2$, we have $\sum_{x\in \mathbb{F}_q}\omega_3^{Tr(ax^d)}=\sum_{x\in \mathbb{F}_q}\omega_3^{Tr(ax^2)}$.

By Lemma \ref{lem_gauss_trace}, it follows that
\begin{eqnarray*}
\sum_{x\in \mathbb{F}_q}\omega_3^{Tr(ax^2)}&=&1+\frac{1}{q-1}\sum_{x\in \mathbb{F}_q^*}\sum_{\chi\in\widehat{\mathbb{F}_{q}^{*}}}G(\chi)\overline{\chi}(ax^2)\\
&=&1+\frac{1}{q-1}\sum_{\chi\in\widehat{\mathbb{F}_{q}^{*}}}G(\chi)\overline{\chi}(a)\sum_{x\in \mathbb{F}_q^*}\overline{\chi}(x^2)\\
&=&1+\sum_{\chi^2=1}G(\chi)\overline{\chi}(a)\\
&=&\eta(a)G(\eta)\\
&=&(-1)^{n+1}\eta(a)i^n3^{n/2}.
\end{eqnarray*}
\done

\begin{theorem}\label{thm_d}
For any $a,\lambda\in \mathbb{F}_q^*$, $\sum_{x\in \mathbb{F}_q}\omega_3^{Tr(ax^d+\lambda x)}=(-1)^{n+1}\eta(a)i^n\omega_3^{g(\lambda)}3^{n/2}$,
where $\eta$ is the quadratic character of $\mathbb{F}_q$, and $$g(\lambda)=\eta(a)\sum_{j: \mathrm{wt}(j)+\mathrm{wt}(-jd)=n+1}\sigma(j)\sigma(-jd)\left(\frac{a}{\lambda^d}\right)^j.$$
\end{theorem}
{\bf Proof. }By Lemma \ref{lem_gauss_trace}, we have
\begin{eqnarray*}
\sum_{x\in \mathbb{F}_q}\omega_3^{Tr(ax^d+\lambda x)}&=&1+\frac{1}{(q-1)^2}\sum_{x\in \mathbb{F}_q^*}\sum_{\chi_1\in\widehat{\mathbb{F}_{q}^{*}}}G(\chi_1)\overline{\chi_1}(ax^d)\sum_{\chi_2\in\widehat{\mathbb{F}_{q}^{*}}}G(\chi_2)\overline{\chi_2}(\lambda x)\\
&=&1+\frac{1}{(q-1)^2}\sum_{\chi_1\in\widehat{\mathbb{F}_{q}^{*}}}\sum_{\chi_2\in\widehat{\mathbb{F}_{q}^{*}}}G(\chi_1)\overline{\chi_1}(a)G(\chi_2)\overline{\chi_2}(\lambda)
\sum_{x\in \mathbb{F}_q^*}\overline{\chi_1}(x^d)\overline{\chi_2}(x)\\
&=&1+\frac{1}{q-1}\sum_{\chi\in\widehat{\mathbb{F}_{q}^{*}}}G(\chi)\overline{\chi}(a)G(\overline{\chi}^d)\chi^d(\lambda)\\
&=&\frac{q}{q-1}+\frac{1}{q-1}\sum_{j=1}^{q-2}G(\chi_\mathfrak{p}^{-j})\chi_\mathfrak{p}^j(a)G(\chi_\mathfrak{p}^{jd})\chi_\mathfrak{p}^{-jd}(\lambda)\\
&=&\frac{q}{q-1}+\frac{-1}{q-1}G(\eta)\eta(a)+\frac{1}{q-1}\sum_{0<j<q-1, j\neq (q-1)/2}G(\chi_\mathfrak{p}^{-j})\chi_\mathfrak{p}^j(a)G(\chi_\mathfrak{p}^{jd})\chi_\mathfrak{p}^{-jd}(\lambda).
\end{eqnarray*}
By Lemma \ref{lem_n} and Theorem \ref{thm_Stickelberger}, we get
$$\frac{\sum_{x\in \mathbb{F}_q}\omega_3^{Tr(ax^d+\lambda x)}}{\pi^n}\in \mathbb{Z}[\omega_3].$$
Because $|\sum_{x\in \mathbb{F}_q}\omega_3^{Tr(ax^d+\lambda x)}|=3^{n/2}$, and $|\pi^n|=3^{n/2}$, it holds that
$$\left|\frac{\sum_{x\in \mathbb{F}_q}\omega_3^{Tr(ax^d+\lambda x)}}{-\eta(a)\pi^n}\right|=1.$$
Thus,
$$\frac{\sum_{x\in \mathbb{F}_q}\omega_3^{Tr(ax^d+\lambda x)}}{-\eta(a)\pi^n}=\pm1, \pm\omega_3, \pm\omega_3^2.$$
One may compute that $$\frac{\sum_{x\in \mathbb{F}_q}\omega_3^{Tr(ax^d+\lambda x)}}{-\eta(a)\pi^n}\equiv1(\mbox{mod }\pi).$$
Hence, $\sum_{x\in \mathbb{F}_q}\omega_3^{Tr(ax^d+\lambda x)}=-\eta(a)\pi^n, -\eta(a)\pi^n\omega_3$, or $-\eta(a)\pi^n\omega_3^2$.
Let $\sum_{x\in \mathbb{F}_q}\omega_3^{Tr(ax^d+\lambda x)}=-\eta(a)\pi^n\omega_3^{h(\lambda)}$, where $h(\lambda)=0, 1, 2$.
Then $\sum_{x\in \mathbb{F}_q}\omega_3^{Tr(ax^d+\lambda x)}=-\eta(a)\pi^n-\eta(a)\pi^{n+1}h(\lambda)+\pi^{n+2}\rho$, where $\rho\in
\mathbb{Z}[\omega_3]$. It follows that
\begin{eqnarray*}
h(\lambda)&=&\frac{\sum_{x\in \mathbb{F}_q}\omega_3^{Tr(ax^d+\lambda x)}+\eta(a)\pi^n}{-\eta(a)\pi^{n+1}}(\mbox{mod }\mathcal{Q})\\
&=&\frac{\eta(a)\pi^n-\eta(a)\omega_3^n\pi^n-\sum_{0<j<q-1, j\neq (q-1)/2}G(\chi_\mathfrak{p}^{-j})\chi_\mathfrak{p}^j(a)G(\chi_\mathfrak{p}^{jd})\chi_\mathfrak{p}^{-jd}(\lambda)}{-\eta(a)\pi^{n+1}}(\mbox{mod }\mathcal{Q})\\
&=&\frac{\omega_3^n\pi^n-\pi^n+\eta(a)\sum_{0<j<q-1, j\neq (q-1)/2}G(\chi_\mathfrak{p}^{-j})\chi_\mathfrak{p}^j(a)G(\chi_\mathfrak{p}^{jd})\chi_\mathfrak{p}^{-jd}(\lambda)}{\pi^{n+1}}(\mbox{mod }\mathcal{Q})\\
&=&n+\eta(a)\frac{\sum_{0<j<q-1}G(\chi_\mathfrak{p}^{-j})\chi_\mathfrak{p}^j(a)G(\chi_\mathfrak{p}^{jd})\chi_\mathfrak{p}^{-jd}(\lambda)}{\pi^{n+1}}(\mbox{mod }\mathcal{Q})\\
&=&\eta(a)\sum_{j: \mathrm{wt}(j)+\mathrm{wt}(-jd)=n+1}\sigma(j)\sigma(-jd)\left(\frac{a}{\lambda^d}\right)^j+n.
\end{eqnarray*}
Because $\pi^n=(-1)^ni^n\omega_3^{2n}3^{n/2}$, we have $\sum_{x\in \mathbb{F}_q}\omega_3^{Tr(ax^d+\lambda x)}=(-1)^{n+1}\eta(a)i^n\omega_3^{h(\lambda)+2n}3^{n/2}$. Let $g(\lambda)=h(\lambda)+2n$. The result follows.
\done

By Theorems \ref{thm_lambda_0} and \ref{thm_d}, we obtain the following main theorem.
\begin{theorem}\label{thm_main}
For any $a\in \mathbb{F}_q^*$ and $\lambda\in \mathbb{F}_q$, $\sum_{x\in \mathbb{F}_q}\omega_3^{Tr(ax^d+\lambda x)}=(-1)^{n+1}\eta(a)i^n\omega_3^{g(\lambda)}3^{n/2}$,
where $\eta$ is the quadratic character of $\mathbb{F}_q$, and $$g(\lambda)=\eta(a)\sum_{j: \mathrm{wt}(j)+\mathrm{wt}(-jd)=n+1}\sigma(j)\sigma(-jd)\left(\frac{a}{\lambda^d}\right)^j,$$
where $a/\lambda^d=0$ if $\lambda=0$.
\end{theorem}

\begin{remark}
If $k=1$, then $d=2$. For this case, in Theorem \ref{thm_main}, one can check that $\{j|0\leq j<3^n-1, \mathrm{wt}(j)+\mathrm{wt}(-jd)=n+1\}=\{\frac{3^n-1}{2}-3^i|i=0, 1, ..., n-1\}$.
For any $j\in\{\frac{3^n-1}{2}-3^i|i=0, 1, ..., n-1\}$, it holds that $\sigma(j)=1$, and $\sigma(-jd)=2$. Thus
\begin{eqnarray*}
g(\lambda)&=&\eta(a)\sum_{j: \mathrm{wt}(j)+\mathrm{wt}(-jd)=n+1}\sigma(j)\sigma(-jd)\left(\frac{a}{\lambda^d}\right)^j\\
&=&2\eta(a)\sum_{i=0}^{n-1}\left(\frac{a}{\lambda^2}\right)^{\frac{3^n-1}{2}-3^i}\\
&=&-\eta(a)a^{\frac{3^n-1}{2}}\sum_{i=0}^{n-1}\left(\frac{a}{\lambda^2}\right)^{-3^i}\\
&=&-Tr\left(\frac{\lambda^2}{a}\right).
\end{eqnarray*}

On the other hand, we have
\begin{eqnarray*}
\sum_{x\in \mathbb{F}_q}\omega_3^{Tr(ax^2+\lambda x)}&=&\sum_{x\in \mathbb{F}_q}\omega_3^{Tr(a(x+\frac{\lambda}{2a})^2-\frac{\lambda^2}{a})}\\
&=&\sum_{x\in \mathbb{F}_q}\omega_3^{Tr(ax^2-\frac{\lambda^2}{a})}\\
&=&(-1)^{n+1}\eta(a)i^n\omega_3^{-Tr(\frac{\lambda^2}{a})}3^{n/2}.
\end{eqnarray*}
Hence, Theorem \ref{thm_main} is correct in this case.
\end{remark}

\begin{theorem}
With notations as in Theorem \ref{thm_main}, if $(-1)^{n+1}\eta(a)i^n=1$, then $Tr(ax^d)$ is a regular bent function.
\end{theorem}
{\bf Proof. }If $(-1)^{n+1}\eta(a)i^n=1$, then $\sum_{x\in \mathbb{F}_q}\omega_3^{Tr(ax^d+\lambda x)}=\omega_3^{g(\lambda)}3^{n/2}$.
Hence, $Tr(ax^d)$ is a regular bent function.
\done

\begin{remark}
1) If $\eta(a)=1$ and $n\equiv 2\bmod 4$, then $(-1)^{n+1}\eta(a)i^n=-i^{n}=1$; 2) if $\eta(a)=-1$ and $n\equiv 0\bmod 4$, then $(-1)^{n+1}\eta(a)i^n=i^{n}=1$.
\end{remark}

%===========================================================================
%===========================================================================
\section{The Case of $n=3t+1$}\label{sec_case1}

In this section, let $k=2t+1$ with $t\geq 2$. Then $d=(3^{2t+1}+1)/2$.

Let $u=\frac{3^n-1}{2}-3^{3t}-3^{2t}-3^t$, $v=\frac{3^n-1}{2}-3^{3t}-3^{2t}+3^{t-1}$, and $w=\frac{3^n-1}{2}-3^{3t}-3^{2t-1}+3^{t-1}$. Firstly, we have the
following lemma.
\begin{lemma}\label{lem_uvw}
$C_u\bigcup C_v\bigcup C_w\subseteq\{j|0\leq j<3^n-1, \mathrm{wt}(j)+\mathrm{wt}(-jd)=n+1\}$. Besides, $\sigma(u)=1$, $\sigma(v)=\sigma(w)=2$,
$\sigma(-ud)=2$, $\sigma(-vd)=1$, and $\sigma(-wd)=2$.
\end{lemma}
{\bf Proof. }

1) One may check that $\mathrm{wt}(u)+\mathrm{wt}(3^ku)=\mathrm{wt}((3^k+1)u)$. We compute $$(3^k+1)u=0\underbrace{11\cdots11}_{t-1}0\underbrace{11\cdots11}_{t-1}01\underbrace{22\cdots22}_{t-1}0\underbrace{11\cdots11}_{t-1}0\underbrace{11\cdots11}_{t}.$$
Then
$$ud=\frac{(3^k+1)u}{2}=\left\{\begin{array}{ll}
                            0\underbrace{0202\cdots02}_{t-2}012\underbrace{0202\cdots02}_{t-2}00\underbrace{22\cdots22}_{t-1}12\underbrace{0202\cdots02}_{t-2}0\underbrace{0202\cdots02}_t, & \mbox{if $t$ is even}; \\
                            0\underbrace{0202\cdots02}_{t-1}0\underbrace{0202\cdots02}_{t-1}00\underbrace{22\cdots22}_{t-1}12\underbrace{0202\cdots02}_{t-3}012\underbrace{0202\cdots02}_{t-1}, & \mbox{if $t$ is odd}.
                          \end{array}\right.$$
Thus, we get
$$ud(\bmod\;3^n-1)=0\underbrace{22\cdots22}_{t-1}1\underbrace{22\cdots22}_{t-1}1\underbrace{22\cdots22}_t,$$
which means that
$$(-ud)(\bmod\;3^n-1)=2\underbrace{00\cdots00}_{t-1}1\underbrace{00\cdots00}_{t-1}1\underbrace{00\cdots00}_t.$$
By Lemma \ref{lem_xy+2}, $2\mathrm{wt}(-ud)=\mathrm{wt}(-(3^k+1)u)+2$. By Lemma \ref{lem_n}, $\mathrm{wt}(u)+\mathrm{wt}(-ud)=n+1$.
Moreover, $\sigma(u)=1$, and $\sigma(-ud)=2$.

2) One may check that $\mathrm{wt}(v)+\mathrm{wt}(3^kv)=\mathrm{wt}((3^k+1)v)+2$. We compute
$$(3^k+1)v=0\underbrace{11\cdots11}_{t-1}0\underbrace{11\cdots11}_{t}\underbrace{22\cdots22}_{t}0\underbrace{11\cdots11}_{t}2\underbrace{11\cdots11}_{t-1}.$$
Then
$$vd=\frac{(3^k+1)v}{2}=\left\{\begin{array}{ll}
0\underbrace{0202\cdots02}_{t-2}012\underbrace{0202\cdots02}_{t-2}0\underbrace{22\cdots22}_{t}12\underbrace{0202\cdots02}_{t-2}022\underbrace{0202\cdots02}_{t-2}, & \mbox{if $t$ is even}; \\
0\underbrace{0202\cdots02}_{t-1}0\underbrace{0202\cdots02}_{t-1}0\underbrace{22\cdots22}_{t}12\underbrace{0202\cdots02}_{t-1}1\underbrace{0202\cdots02}_{t-1}, & \mbox{if $t$ is odd}.
                          \end{array}\right.$$
Thus, we get
$$vd(\bmod\;3^n-1)=\underbrace{22\cdots22}_{t}1\underbrace{22\cdots22}_{t}1\underbrace{22\cdots22}_{t-1},$$
which means that
$$(-vd)(\bmod\;3^n-1)=\underbrace{00\cdots00}_{t}1\underbrace{00\cdots00}_{t}1\underbrace{00\cdots00}_{t-1}.$$
By Lemma \ref{lem_xy+2}, $2\mathrm{wt}(-vd)=\mathrm{wt}(-(3^k+1)v)$. By Lemma \ref{lem_n}, $\mathrm{wt}(v)+\mathrm{wt}(-vd)=n+1$.
Moreover, $\sigma(v)=2$, and $\sigma(-vd)=1$.

3) One may check that $\mathrm{wt}(w)+\mathrm{wt}(3^kw)=\mathrm{wt}((3^k+1)w)$. We compute
$$(3^k+1)w=0\underbrace{11\cdots11}_{t}0\underbrace{11\cdots11}_{t-1}\underbrace{22\cdots22}_{t}10\underbrace{11\cdots11}_{t-1}2\underbrace{11\cdots11}_{t-1}.$$
$$wd=\frac{(3^k+1)w}{2}=\left\{\begin{array}{ll}
0\underbrace{0202\cdots02}_{t}0\underbrace{0202\cdots02}_{t-2}0\underbrace{22\cdots22}_{t+1}0\underbrace{0202\cdots02}_{t-2}022\underbrace{0202\cdots02}_{t-2}, & \mbox{if $t$ is even}; \\
0\underbrace{0202\cdots02}_{t-1}012\underbrace{0202\cdots02}_{t-3}0\underbrace{22\cdots22}_{t+1}0\underbrace{0202\cdots02}_{t-1}1\underbrace{0202\cdots02}_{t-1}, & \mbox{if $t$ is odd}.
                          \end{array}\right.$$
Thus, we get
$$wd(\bmod\;3^n-1)=\underbrace{22\cdots22}_{t+1}0\underbrace{22\cdots22}_{2t-1},$$
which means that
$$(-wd)(\bmod\;3^n-1)=\underbrace{00\cdots00}_{t+1}2\underbrace{00\cdots00}_{2t-1}.$$
By Lemma \ref{lem_xy+2}, $2\mathrm{wt}(-wd)=\mathrm{wt}(-(3^k+1)w)+2$. By Lemma \ref{lem_n}, $\mathrm{wt}(w)+\mathrm{wt}(-wd)=n+1$.
Moreover, $\sigma(w)=2$, and $\sigma(-wd)=2$.
\done

\begin{lemma}\label{lem_1_1}
For any $0<j<3^n-1$ with $j\neq (3^n-1)/2$, let $h=h_{n-1}h_{n-2}\cdots h_1h_0=\overline{(3^k+1)j}$, and $u=u_{n-1}u_{n-2}\cdots u_1u_0=\overline{jd}$. If $\mathrm{wt}(j)+\mathrm{wt}(-jd)=n+1$ and $\mathrm{wt}(j)+\mathrm{wt}(3^kj)=\mathrm{wt}((3^k+1)j)$, then there exists $0\leq a<n$ such that $h_a=h_{a+1}=1$, and for any $i\neq a, a+1$, $h_i=0$ or 2. Besides, if $a=0$, then $u=(\frac{h}{2}+\frac{3^n-1}{2})(\bmod\;3^n-1)$.
\end{lemma}
{\bf Proof. }By Lemma \ref{lem_n}, $2\mathrm{wt}(-u)=\mathrm{wt}(-2u)+2$.
Suppose that $h_i=0$ or 2 for any $0\leq i<n$. We consider two cases.

1) $u=\frac{h}{2}$. Because $0<j<3^n-1$ and $j\neq (3^n-1)/2$, there exists $0\leq a<n$ such that $h_a=0$ and $h_{a+1}=2$.
It follows that $u_a=0$ and $u_{a+1}=1$. Hence, $2\mathrm{wt}(-u)>\mathrm{wt}(-2u)+2$.

2) $u=(\frac{h}{2}+\frac{3^n-1}{2})(\bmod\;3^n-1)$. Then $u_i=1$ or 2 for any $0\leq i<n$. Hence, $2\mathrm{wt}(-u)=\mathrm{wt}(-2u)$.

Because $h$ is even, there are even number of 1's in the ternary representation of $h$. Suppose that there are four 1's in the ternary representation of $h$ (the case of more 1's is similar), i.e., there exist $0\leq a<b<c<d<n$ such that $h_a=h_b=h_c=h_d=1$. We also consider two cases.

1) $u=\frac{h}{2}$. Then $u_b=u_d=0$. Hence, $2\mathrm{wt}(-u)>\mathrm{wt}(-2u)+2$.

2) $u=(\frac{h}{2}+\frac{3^n-1}{2})(\bmod\;3^n-1)$. Then $u_a=u_c=0$. Hence, $2\mathrm{wt}(-u)>\mathrm{wt}(-2u)+2$.

Hence there are only two 1's in the ternary representation of $h$, i.e., there exist $0\leq a<b<n$ such that $h_a=h_b=1$. Suppose that $b\neq a+1$.
If $u=\frac{h}{2}$, then $u_b=0$, $u_{b+1}=0$ or 1, which means that $2\mathrm{wt}(-u)>\mathrm{wt}(-2u)+2$.
If $u=(\frac{h}{2}+\frac{3^n-1}{2})(\bmod\;3^n-1)$, then $u_b=0$, $u_{b+1}=0$ or 1, which means that $2\mathrm{wt}(-u)>\mathrm{wt}(-2u)+2$. We get a contradiction. Thus, $b=a+1$.

If $a=0$ and $u=\frac{h}{2}$, then $u_1=0$, $u_2=0$ or 1, which means that $2\mathrm{wt}(-u)>\mathrm{wt}(-2u)+2$. We get a contradiction. Hence, $u=(\frac{h}{2}+\frac{3^n-1}{2})(\bmod\;3^n-1)$. \done

\begin{lemma}\label{lem_h/2}
For any $0<j<3^n-1$, let $h=\overline{(3^k+1)j}$, and $u=\overline{jd}$. If $\overline{3^kj}+j<3^n-1$, then $u=(\frac{h}{2}+\frac{3^n-1}{2})(\bmod\;3^n-1)$ if and only if $\sum_{i=n-k}^{n-1}j_i(\bmod\;2)=1$, where $j_{n-1}j_{n-2}\cdots j_1j_0$ is the ternary representation of $j$.
\end{lemma}
{\bf Proof. }Because $\overline{3^kj}+j<3^n-1$, we have $h=\overline{3^kj}+j$.
It holds that $$\overline{3^kj}=\sum_{i=0}^{n-k-1}j_i3^{i+k}+\sum_{i=n-k}^{n-1}j_i3^{i+k-n}.$$
Hence, $$3^kj-\overline{3^kj}=\sum_{i=n-k}^{n-1}j_i3^{i+k}-\sum_{i=n-k}^{n-1}j_i3^{i+k-n},$$
which means that $3^kj=\overline{3^kj}+(3^n-1)\sum_{i=n-k}^{n-1}j_i3^{i+k-n}$.
Thus, $$(3^k+1)j=\overline{3^kj}+j+(3^n-1)\sum_{i=n-k}^{n-1}j_i3^{i+k-n}=h+(3^n-1)\sum_{i=n-k}^{n-1}j_i3^{i+k-n}.$$
Because $$\sum_{i=n-k}^{n-1}j_i3^{i+k-n}(\bmod\;2)=\sum_{i=n-k}^{n-1}j_i(\bmod\;2),$$ we obtain $u=\overline{(3^k+1)j/2}=(\frac{h}{2}+\frac{3^n-1}{2})(\bmod\;3^n-1)$ if and only if $\sum_{i=n-k}^{n-1}j_i(\bmod\;2)=1$.\done

\begin{lemma}\label{lem_uw}
For any $0<j<3^n-1$, if $\mathrm{wt}(j)+\mathrm{wt}(-jd)=n+1$ and $\mathrm{wt}(j)+\mathrm{wt}(3^kj)=\mathrm{wt}((3^k+1)j)$, then
$j\in C_u\bigcup C_w$.
\end{lemma}
{\bf Proof. }Let $h=h_{n-1}h_{n-2}\cdots h_1h_0=\overline{(3^k+1)j}$, and $j=j_{n-1}j_{n-2}\cdots j_1j_0$. By Lemma \ref{lem_1_1}, there exists $0\leq a<n$ such that $h_a=h_{a+1}=1$.
Without loss of generality, we can assume that $a=0$, i.e., $h_0=h_1=1$. We consider four cases below.

1) $j_0=j_1=1$. Then $j_{-k}=j_{-k+1}=0$. Because $j_0+j_k=2$, it holds that $j_k=1$. Similarly, $j_{2k}=1$. However, $j_{-k+1}=j_{2k}$.
It is a contradiction, and this case is impossible.

2) $j_0=j_1=0$. Then $j_{-k}=j_{-k+1}=1$. Because $j_{-k+1}+j_{-2k+1}=2$, it holds that $j_{-2k+1}=1$. Similarly, $j_{-3k+1}=1$. However, $j_{-3k+1}=j_0$. It is a contradiction, and this case is impossible.

3) $j_0=0, j_1=1$. Then $j_{-k}=1, j_{-k+1}=0$. It follows that $j_{-ik}=1$ for any $1\leq i\leq n-3$. Because $j_0+j_k=0$ or 2, it holds that $j_k=0$ or 2.
If $j_k=0$, then $j\in C_v$; otherwise, $j\in C_w$.

4) $j_0=1, j_1=0$. Then $j_{-k}=0, j_{-k+1}=1$. It follows that $j_{-2k+1}=j_{-3k+1}=1$. Note that $j_{-3k+1}=j_0$.
If there exists $0<i<n$ with $i\neq -k+1, -2k+1$ such that $j_i=1$, then we can get $j_1=1$ or $j_{-k}=1$. It is a contradiction. Thus, $j_i=0$ or 2 for any $i\neq 1, -k$. However, in this case, $\sum_{i=n-k}^{n-1}j_i(\bmod\;2)=0$. By Lemma \ref{lem_h/2}, $u=\frac{h}{2}$.
By Lemma \ref{lem_1_1}, it is a contradiction again. Hence, this case is impossible.\done

\begin{lemma}\label{lem_1_2}
For any $0<j<3^n-1$ with $j\neq (3^n-1)/2$, let $h=h_{n-1}h_{n-2}\cdots h_1h_0=\overline{(3^k+1)j}$, and $u=u_{n-1}u_{n-2}\cdots u_1u_0=\overline{jd}$. If $\mathrm{wt}(j)+\mathrm{wt}(-jd)=n+1$ and $\mathrm{wt}(j)+\mathrm{wt}(3^kj)=\mathrm{wt}((3^k+1)j)+2$, then
$h_i=0$ or 2 for any $0\leq i<n$, and there exists $0\leq a<n$ such that $\{j_a, j_{-k+a}\}=\{2, 1\}$ and $\{j_{a+1}, j_{-k+a+1}\}=\{0, 1\}$, where $j_{n-1}j_{n-2}\cdots j_1j_0$ is the ternary representation of $j$. Besides, if $a=0$, then $u=(\frac{h}{2}+\frac{3^n-1}{2})(\bmod\;3^n-1)$.

\end{lemma}
{\bf Proof. }
If $\mathrm{wt}(j)+\mathrm{wt}(-jd)=n+1$ and $\mathrm{wt}(j)+\mathrm{wt}(3^kj)=\mathrm{wt}((3^k+1)j)+2$,
then $2\mathrm{wt}(-u)=\mathrm{wt}(-2u)$ by Lemma \ref{lem_n}. Suppose that there exists $0\leq a<n$ such that $h_a=1$. Similar to the proof of Lemma \ref{lem_1_1}, we can show that $2\mathrm{wt}(-u)>\mathrm{wt}(-2u)$. Hence, $h_i=0$ or 2 for any $0\leq i<n$.

Because $\mathrm{wt}(j)+\mathrm{wt}(3^kj)=\mathrm{wt}((3^k+1)j)+2$, by Lemma \ref{lem_xy+2}, there exists $0\leq a<n$ such that $j_a+j_{-k+a}\geq3$ and $j_{a+1}+j_{-k+a+1}\leq1$. Because $h_a=0$ or 2, and $h_{a+1}=0$ or 2, we have $j_a+j_{-k+a}=3$ and $j_{a+1}+j_{-k+a+1}=1$. Therefore, $\{j_a, j_{-k+a}\}=\{2, 1\}$, and $\{j_{a+1}, j_{-k+a+1}\}=\{0, 1\}$.

If $a=0$, then $h_0=0$, and $h_1=2$. If $u=h/2$, then $2\mathrm{wt}(-u)>\mathrm{wt}(-2u)$. It is a contradiction. Therefore, $u=(\frac{h}{2}+\frac{3^n-1}{2})(\bmod\;3^n-1)$. \done

\begin{lemma}\label{lem_v}
For any $0<j<3^n-1$, if $\mathrm{wt}(j)+\mathrm{wt}(-jd)=n+1$ and $\mathrm{wt}(j)+\mathrm{wt}(3^kj)=\mathrm{wt}((3^k+1)j)+2$, then
$j\in C_v$.
\end{lemma}
{\bf Proof. }Let $j_{n-1}j_{n-2}\cdots j_1j_0$ be the ternary representation of $j$. Let $h=h_{n-1}h_{n-2}\cdots h_1h_0=\overline{(3^k+1)j}$, and $u=u_{n-1}u_{n-2}\cdots u_1u_0=\overline{jd}$. By Lemma \ref{lem_1_2}, there exists $0\leq a<n$ such that $\{j_a, j_{-k+a}\}=\{2, 1\}$ and $\{j_{a+1}, j_{-k+a+1}\}=\{0, 1\}$, and $h_i=0$ or 2 for any $0\leq i<n$. Without loss of generality, we can assume that $a=0$.
In the following, we consider four cases.

1) $j_0=1, j_1=0$. Then $j_{-k}=2, j_{-k+1}=1$. Because $j_{-k+1}+j_{-2k+1}=2$, we have $j_{-2k+1}=1$. Note that $j_{-3k+1}=j_0$.
For any $1<i<n$ with $i\neq -k, -k+1, -2k+1$, if $j_i=1$, then we can get that $j_1=1$ or $j_{-k}=1$. It is a contradiction.
Hence, $j_i=0$ or 2 for any $1<i<n$ with $i\neq -k, -k+1, -2k+1$. However, in this case, $\sum_{i=n-k}^{n-1}j_i(\bmod\;2)=0$. By Lemma \ref{lem_h/2}, $u=\frac{h}{2}$.
By Lemma \ref{lem_1_2}, it is a contradiction again. Therefore, this case is impossible.

2) $j_0=1, j_1=1$. Then $j_{-k}=2, j_{-k+1}=0$. Similarly, we can get that $j_k=j_{2k}=1$. Note that $2k\equiv -k+1(\bmod\;n)$. Thus, $j_{-k+1}=1$.
It is a contradiction. This case is impossible.

3) $j_0=2, j_1=0$. Then $j_{-k}=1, j_{-k+1}=1$. It follows that $j_{-ik}=1$ for any $1\leq i\leq n-3$. Thus $j_{-(n-3)k}=j_{-nk+3k}=j_{3k}=1$,
which means that $j_1=1$. It is a contradiction. This case is impossible.

4) $j_0=2, j_1=1$. Then $j_{-k}=1, j_{-k+1}=0$. It follows that $j_{-ik}=1$ for any $1\leq i\leq n-3$.
Because $j_0+j_k=2$, it holds that $j_k=2$. Thus, $j\in C_v$.
\done

\begin{theorem}
For any $a\in \mathbb{F}_q^*$ and $\lambda\in \mathbb{F}_q$, $\sum_{x\in \mathbb{F}_q}\omega_3^{Tr(ax^d+\lambda x)}=(-1)^{n+1}\eta(a)i^n\omega_3^{g(\lambda)}3^{n/2}$,
where $\eta$ is the quadratic character of $\mathbb{F}_q$, and
$$g(\lambda)=Tr\left(-\frac{\lambda^{3^{2t+1}+3^{t+1}+2}}{a^{3^{2t+1}+3^{t+1}+1}}-\frac{\lambda^{3^{2t}+1}}{a^{-3^{2t}+3^{t}+1}}+\frac{\lambda^{2}}{a^{-3^{2t+1}+3^{t+1}+1}}\right).$$
\end{theorem}
{\bf Proof. }By Theorem \ref{thm_main}, we have $\sum_{x\in \mathbb{F}_q}\omega_3^{Tr(ax^d+\lambda x)}=(-1)^{n+1}\eta(a)i^n\omega_3^{g(\lambda)}3^{n/2}$, where
$$g(\lambda)=\eta(a)\sum_{j: \mathrm{wt}(j)+\mathrm{wt}(-jd)=n+1}\sigma(j)\sigma(-jd)\left(\frac{a}{\lambda^d}\right)^j.$$
By Lemmas \ref{lem_uvw}, \ref{lem_uw}, and \ref{lem_v}, we have
$$C_u\bigcup C_v\bigcup C_w=\{j|0\leq j<3^n-1, \mathrm{wt}(j)+\mathrm{wt}(-jd)=n+1\}.$$
Thus, by Lemma \ref{lem_uvw} and its proof, it holds that
\begin{eqnarray*}
g(\lambda)&=&\eta(a)Tr\left(-\left(\frac{a}{\lambda^d}\right)^u-\left(\frac{a}{\lambda^d}\right)^v+\left(\frac{a}{\lambda^d}\right)^w\right)\\
&=&\eta(a)Tr(-a^u\lambda^{-ud}-a^v\lambda^{-vd}+a^w\lambda^{-wd})\\
&=&\eta(a)Tr(-a^u\lambda^{2\cdot3^{3t}+3^{2t}+3^t}-a^v\lambda^{3^{2t}+3^{t-1}}+a^w\lambda^{2\cdot 3^{2t-1}})\\
&=&\eta(a)Tr\left(-\frac{a^{\frac{3^n-1}{2}}\lambda^{2\cdot3^{3t}+3^{2t}+3^t}}{a^{3^{3t}+3^{2t}+3^t}}-\frac{a^{\frac{3^n-1}{2}}\lambda^{3^{2t}+3^{t-1}}}{a^{3^{3t}+3^{2t}-3^{t-1}}}+\frac{a^{\frac{3^n-1}{2}}\lambda^{2\cdot 3^{2t-1}}}{a^{3^{3t}+3^{2t-1}-3^{t-1}}}\right)\\
&=&Tr\left(-\frac{\lambda^{2\cdot3^{3t}+3^{2t}+3^t}}{a^{3^{3t}+3^{2t}+3^t}}-\frac{\lambda^{3^{2t}+3^{t-1}}}{a^{3^{3t}+3^{2t}-3^{t-1}}}+\frac{\lambda^{2\cdot 3^{2t-1}}}{a^{3^{3t}+3^{2t-1}-3^{t-1}}}\right)\\
&=&Tr\left(-\frac{\lambda^{3^{2t+1}+3^{t+1}+2}}{a^{3^{2t+1}+3^{t+1}+1}}-\frac{\lambda^{3^{2t}+1}}{a^{-3^{2t}+3^{t}+1}}+\frac{\lambda^{2}}{a^{-3^{2t+1}+3^{t+1}+1}}\right).
\end{eqnarray*}\done

\begin{example}
For the case of $\mathbb{F}_{3^7}$, the dual function is given by
$$g(\lambda)=Tr\left(-\frac{\lambda^{272}}{a^{271}}-\frac{\lambda^{82}}{a^{-71}}+\frac{\lambda^{2}}{a^{-215}}\right)=Tr(-a^{1915}\lambda^{272}-a^{71}\lambda^{82}+a^{215}\lambda^2).$$
For the case of $\mathbb{F}_{3^{10}}$, the dual function is given by
$$g(\lambda)=Tr\left(-\frac{\lambda^{2270}}{a^{2269}}-\frac{\lambda^{730}}{a^{-701}}+\frac{\lambda^{2}}{a^{-2105}}\right)=Tr(-a^{56779}\lambda^{2270}-a^{701}\lambda^{730}+a^{2105}\lambda^{2}).$$
\end{example}

%===========================================================================
%===========================================================================
\section{The Case of $n=3t+2$}\label{sec_case2}

In this section, let $k=2t+1$ with $t\geq 2$. Then $d=(3^{2t+1}+1)/2$.

Let $u=\frac{3^n-1}{2}-3^{3t+1}-3^{2t+1}-3^t$, $v=\frac{3^n-1}{2}-3^{3t+1}-3^{2t+1}+3^{t}$, and $w=\frac{3^n-1}{2}-3^{3t+1}-3^{2t}+3^{t-1}$.
Then we have the following lemma.

\begin{lemma}\label{lem_uvw_2}
$C_u\bigcup C_v\bigcup C_w\subseteq\{j|0\leq j<3^n-1, \mathrm{wt}(j)+\mathrm{wt}(-jd)=n+1\}$. Besides, $\sigma(u)=1$, $\sigma(v)=\sigma(w)=2$,
$\sigma(-ud)=\sigma(-vd)=2$, and $\sigma(-wd)=1$.
\end{lemma}
{\bf Proof. }

1) One may check that $\mathrm{wt}(u)+\mathrm{wt}(3^ku)=\mathrm{wt}((3^k+1)u)$. We compute $$(3^k+1)u=0\underbrace{11\cdots11}_{t-1}0\underbrace{11\cdots11}_{t}0\underbrace{22\cdots22}_{t-1}\underbrace{11\cdots11}_{t+1}0\underbrace{11\cdots11}_{t}.$$
Then
$$ud=\frac{(3^k+1)u}{2}=\left\{\begin{array}{ll}
                            0\underbrace{02\cdots02}_{t-2}012\underbrace{02\cdots02}_{t-2}012\underbrace{22\cdots22}_{t-1}\underbrace{02\cdots02}_{t}0\underbrace{02\cdots02}_{t}, & \mbox{if $t$ is even}; \\
                            0\underbrace{02\cdots02}_{t-1}0\underbrace{02\cdots02}_{t-1}012\underbrace{22\cdots22}_{t-1}\underbrace{02\cdots02}_{t-1}012\underbrace{02\cdots02}_{t-1}, & \mbox{if $t$ is odd}.
                          \end{array}\right.$$
Thus, we get
$$ud(\bmod\;3^n-1)=1\underbrace{22\cdots22}_{t}0\underbrace{22\cdots22}_{t-1}1\underbrace{22\cdots22}_{t},$$
which means that
$$(-ud)(\bmod\;3^n-1)=1\underbrace{00\cdots00}_{t}2\underbrace{00\cdots00}_{t-1}1\underbrace{00\cdots00}_{t}.$$
By Lemma \ref{lem_xy+2}, $2\mathrm{wt}(-ud)=\mathrm{wt}(-(3^k+1)u)+2$. By Lemma \ref{lem_n}, $\mathrm{wt}(u)+\mathrm{wt}(-ud)=n+1$.
Moreover, $\sigma(u)=1$, and $\sigma(-ud)=2$.

2) One may check that $\mathrm{wt}(v)+\mathrm{wt}(3^kv)=\mathrm{wt}((3^k+1)v)$. We compute
$$(3^k+1)v=0\underbrace{11\cdots11}_{t-1}0\underbrace{11\cdots11}_{t}\underbrace{22\cdots22}_{t}\underbrace{11\cdots11}_{t+1}2\underbrace{11\cdots11}_{t}.$$
Then
$$vd=\frac{(3^k+1)v}{2}=\left\{\begin{array}{ll}
0\underbrace{02\cdots02}_{t-2}012\underbrace{02\cdots02}_{t-2}0\underbrace{22\cdots22}_{t+1}\underbrace{02\cdots02}_{t}1\underbrace{02\cdots02}_{t}, & \mbox{if $t$ is even}; \\
0\underbrace{02\cdots02}_{t-1}0\underbrace{02\cdots02}_{t-1}0\underbrace{22\cdots22}_{t+1}\underbrace{02\cdots02}_{t-1}022\underbrace{02\cdots02}_{t-1}, & \mbox{if $t$ is odd}.
                          \end{array}\right.$$
Thus, we get
$$vd(\bmod\;3^n-1)=\underbrace{22\cdots22}_{t+1}0\underbrace{22\cdots22}_{2t},$$
which means that
$$(-vd)(\bmod\;3^n-1)=\underbrace{00\cdots00}_{t+1}2\underbrace{00\cdots00}_{2t}.$$
By Lemma \ref{lem_xy+2}, $2\mathrm{wt}(-vd)=\mathrm{wt}(-(3^k+1)v)+2$. By Lemma \ref{lem_n}, $\mathrm{wt}(v)+\mathrm{wt}(-vd)=n+1$.
Moreover, $\sigma(v)=2$, and $\sigma(-vd)=2$.

3) One may check that $\mathrm{wt}(w)+\mathrm{wt}(3^kw)=\mathrm{wt}((3^k+1)w)+2$. We compute
$$(3^k+1)w=0\underbrace{11\cdots11}_{t}0\underbrace{11\cdots11}_{t-1}20\underbrace{22\cdots22}_{t-1}0\underbrace{11\cdots11}_{t}2\underbrace{11\cdots11}_{t-1}.$$
Then
$$wd=\frac{(3^k+1)w}{2}=\left\{\begin{array}{ll}
0\underbrace{02\cdots02}_{t}0\underbrace{02\cdots02}_{t}12\underbrace{22\cdots22}_{t-2}12\underbrace{02\cdots02}_{t-2}022\underbrace{02\cdots02}_{t-2}, & \mbox{if $t$ is even}; \\
0\underbrace{02\cdots02}_{t-1}012\underbrace{02\cdots02}_{t-1}12\underbrace{22\cdots22}_{t-2}12\underbrace{02\cdots02}_{t-1}1\underbrace{02\cdots02}_{t-1}, & \mbox{if $t$ is odd}.
                          \end{array}\right.$$
Thus, we get
$$wd(\bmod\;3^n-1)=21\underbrace{22\cdots22}_{t-1}1\underbrace{22\cdots22}_{2t},$$
which means that
$$(-wd)(\bmod\;3^n-1)=01\underbrace{00\cdots00}_{t-1}1\underbrace{00\cdots00}_{2t}.$$
By Lemma \ref{lem_xy+2}, $2\mathrm{wt}(-wd)=\mathrm{wt}(-(3^k+1)w)$. By Lemma \ref{lem_n}, $\mathrm{wt}(w)+\mathrm{wt}(-wd)=n+1$.
Moreover, $\sigma(w)=2$, and $\sigma(-wd)=1$. \done

\begin{lemma}\label{lem_uv_2}
For any $0<j<3^n-1$, if $\mathrm{wt}(j)+\mathrm{wt}(-jd)=n+1$ and $\mathrm{wt}(j)+\mathrm{wt}(3^kj)=\mathrm{wt}((3^k+1)j)$, then
$j\in C_u\bigcup C_v$.
\end{lemma}
{\bf Proof. }Let $j=j_{n-1}j_{n-2}\cdots j_1j_0$, and $h=h_{n-1}h_{n-2}\cdots h_1h_0=\overline{(3^k+1)j}$.
By Lemma \ref{lem_1_1}, there exists $0\leq a<n$ such that $h_a=h_{a+1}=1$. Without loss of generality, we can assume that $a=0$, i.e., $h_0=h_1=1$.
There are four cases to be considered.

1) $j_0=j_1=1$. Then $j_{-k}=j_{-k+1}=0$. Because $j_0+j_k=2$, it holds that $j_k=1$. Similarly, we can get that $j_{ik}=1$ for $0\leq i\leq n-2$.
Because $j_{(n-2)k}=1$, it is a contradiction. This case is impossible.

2) $j_0=j_1=0$. Then $j_{-k}=j_{-k+1}=1$. Similarly, we can get that $j_{-k}=j_{-2k}=j_{-3k}=1$. However, $j_{-3k}=j_1$. It is a contradiction.
This case is impossible.

3) $j_0=0, j_1=1$. Then $j_{-k}=1, j_{-k+1}=0$. It follows that $j_{-2k}=1$. Note that $j_{-3k}=j_1$. For any $1<i<n$ with $i\neq-k, -2k, -k+1$, if
$j_i=1$, then we can get that $j_0=1$ or $j_{-k+1}=1$. It is a contradiction. Hence, $j_i=0$ or 2 for any $1<i<n$ with $i\neq-k, -2k, -k+1$.
However, in this case, $\sum_{i=n-k}^{n-1}j_i(\bmod\;2)=0$. By Lemma \ref{lem_h/2}, $u=\frac{h}{2}$.
By Lemma \ref{lem_1_1}, it is a contradiction again. Therefore, this case is impossible.

4) $j_0=1, j_1=0$. Then $j_{-k}=0, j_{-k+1}=1$. It follows that $j_{ik}=0$ for any $0\leq i\leq n-4$. Note that $j_{(n-3)k}=j_1$.
Because $j_{(n-2)k}+j_{-k}=0$ or 2, it holds that $j_{(n-2)k}=0$ or 2. If $j_{(n-2)k}=0$, then $j\in C_u$; otherwise, $j\in C_v$.
\done

\begin{lemma}\label{lem_w_2}
For any $0<j<3^n-1$, if $\mathrm{wt}(j)+\mathrm{wt}(-jd)=n+1$ and $\mathrm{wt}(j)+\mathrm{wt}(3^kj)=\mathrm{wt}((3^k+1)j)+2$, then
$j\in C_w$.
\end{lemma}
{\bf Proof. }Let $j=j_{n-1}j_{n-2}\cdots j_1j_0$, $h=h_{n-1}h_{n-2}\cdots h_1h_0=\overline{(3^k+1)j}$, and $u=u_{n-1}u_{n-2}\cdots u_1u_0=\overline{jd}$.
By Lemma \ref{lem_1_2}, there exists $0\leq a<n$ such that $\{j_a, j_{-k+a}\}=\{2, 1\}$, $\{j_{a+1}, j_{-k+a+1}\}=\{0, 1\}$, and $h_i=0$ or 2 for any $0\leq i<n$. Without loss of generality, we can assume that $a=0$. There are four cases to be considered.

1) $j_0=1, j_1=0$. Then $j_{-k}=2, j_{-k+1}=1$. It follows that $j_{ik}=1$ for $0\leq i\leq n-4$.
Note that $j_{(n-4)k}=j_{-k+1}$ and $j_{(n-3)k}=j_1$. Because $j_{-k}+j_{-2k}=0$ or 2, it holds that $j_{-2k}=0$. Then one may check that $j\in C_w$.

2) $j_0=1, j_1=1$. Then $j_{-k}=2, j_{-k+1}=0$. It follows that $j_{ik}=1$ for $0\leq i\leq n-4$. However, $j_{(n-4)k}=j_{-k+1}$.
It is a contradiction. This case is impossible.

3) $j_0=2, j_1=0$. Then $j_{-k}=1, j_{-k+1}=1$. It follows that $j_{-k}=j_{-2k}=j_{-3k}=1$. Note that $j_{-3k}=j_1$.
It is a contradiction. This case is impossible.

4) $j_0=2, j_1=1$. Then $j_{-k}=1, j_{-k+1}=0$. It follows that $j_{-k}=j_{-2k}=j_{-3k}=1$. Note that $j_{-3k}=j_1$.
For any $1<i<n$ with $i\neq-k, -2k, -k+1$, if $j_i=1$, then we can get that $j_0=1$ or $j_{-k+1}=1$. It is a contradiction.
Hence, $j_i=0$ or 2 for any $1<i<n$ with $i\neq-k, -2k, -k+1$. However, in this case, $\sum_{i=n-k}^{n-1}j_i(\bmod\;2)=0$. By Lemma \ref{lem_h/2}, $u=\frac{h}{2}$.
By Lemma \ref{lem_1_2}, it is a contradiction again. Therefore, this case is impossible. \done

\begin{theorem}
For any $a\in \mathbb{F}_q^*$ and $\lambda\in \mathbb{F}_q$, $\sum_{x\in \mathbb{F}_q}\omega_3^{Tr(ax^d+\lambda x)}=(-1)^{n+1}\eta(a)i^n\omega_3^{g(\lambda)}3^{n/2}$,
where $\eta$ is the quadratic character of $\mathbb{F}_q$, and
$$g(\lambda)=Tr\left(-\frac{\lambda^{3^{2t+2}+1}}{a^{3^{2t+2}-3^{t+1}+3}}-\frac{\lambda^{2\cdot3^{2t+1}+3^{t+1}+1}}{a^{3^{2t+2}+3^{t+1}+1}}+\frac{\lambda^2}{a^{-3^{2t+2}+3^{t+1}+3}}\right).$$
\end{theorem}
{\bf Proof. }By Theorem \ref{thm_main}, we have $\sum_{x\in \mathbb{F}_q}\omega_3^{Tr(ax^d+\lambda x)}=(-1)^{n+1}\eta(a)i^n\omega_3^{g(\lambda)}3^{n/2}$, where
$$g(\lambda)=\eta(a)\sum_{j: \mathrm{wt}(j)+\mathrm{wt}(-jd)=n+1}\sigma(j)\sigma(-jd)\left(\frac{a}{\lambda^d}\right)^j.$$
By Lemmas \ref{lem_uvw_2}, \ref{lem_uv_2}, and \ref{lem_w_2}, we have
$$C_u\bigcup C_v\bigcup C_w=\{j|0\leq j<3^n-1, \mathrm{wt}(j)+\mathrm{wt}(-jd)=n+1\}.$$
Thus, by Lemma \ref{lem_uvw_2} and its proof, it holds that
\begin{eqnarray*}
g(\lambda)&=&\eta(a)Tr\left(-\left(\frac{a}{\lambda^d}\right)^u+\left(\frac{a}{\lambda^d}\right)^v-\left(\frac{a}{\lambda^d}\right)^w\right)\\
&=&\eta(a)Tr(-a^u\lambda^{-ud}+a^v\lambda^{-vd}-a^w\lambda^{-wd})\\
&=&\eta(a)Tr(-a^u\lambda^{3^{3t+1}+2\cdot3^{2t}+3^t}+a^v\lambda^{2\cdot3^{2t}}-a^w\lambda^{3^{3t}+3^{2t}})\\
&=&\eta(a)Tr\left(-\frac{a^{\frac{3^n-1}{2}}\lambda^{3^{3t+1}+2\cdot3^{2t}+3^t}}{a^{3^{3t+1}+3^{2t+1}+3^t}}+\frac{a^{\frac{3^n-1}{2}}\lambda^{2\cdot3^{2t}}}{a^{3^{3t+1}+3^{2t+1}-3^{t}}}
-\frac{a^{\frac{3^n-1}{2}}\lambda^{3^{3t}+3^{2t}}}{a^{3^{3t+1}+3^{2t}-3^{t-1}}}\right)\\
&=&Tr\left(-\frac{\lambda^{3^{3t+1}+2\cdot3^{2t}+3^t}}{a^{3^{3t+1}+3^{2t+1}+3^t}}+\frac{\lambda^{2\cdot3^{2t}}}{a^{3^{3t+1}+3^{2t+1}-3^{t}}}
-\frac{\lambda^{3^{3t}+3^{2t}}}{a^{3^{3t+1}+3^{2t}-3^{t-1}}}\right)\\
&=&Tr\left(-\frac{\lambda^{3^{2t+2}+1}}{a^{3^{2t+2}-3^{t+1}+3}}-\frac{\lambda^{2\cdot3^{2t+1}+3^{t+1}+1}}{a^{3^{2t+2}+3^{t+1}+1}}+\frac{\lambda^2}{a^{-3^{2t+2}+3^{t+1}+3}}\right).
\end{eqnarray*}\done

\begin{example}
For the case of $\mathbb{F}_{3^8}$, the dual function is given by
$$g(\lambda)=Tr\left(-\frac{\lambda^{730}}{a^{705}}-\frac{\lambda^{514}}{a^{757}}+\frac{\lambda^2}{a^{-699}}\right)=Tr(-a^{5855}\lambda^{730}-a^{5803}\lambda^{514}+a^{699}\lambda^2).$$
For the case of $\mathbb{F}_{3^{11}}$, the dual function is given by
$$g(\lambda)=Tr\left(-\frac{\lambda^{6562}}{a^{6483}}-\frac{\lambda^{4456}}{a^{6643}}+\frac{\lambda^2}{a^{-6477}}\right)=Tr(-a^{170663}\lambda^{6562}-a^{170503}\lambda^{4456}+a^{6477}\lambda^2).$$
\end{example}

%===========================================================================
%===========================================================================
\section{Concluding Remarks}\label{sec_con}

In this paper, we investigate the dual function of the Coulter-Matthews bent function. Via Stickelberger's theorem
and the Teichm\"{u}ller character, we find an universal formula for the dual function. For two special cases,
using new combinatorial methods we develop in this paper, we determine the formula explicitly which has only three terms.
From the viewpoint of both bent functions and character sums, the results of this paper are very interesting. Moreover,
the findings of this paper can be applied to the correlation distribution of the Coulter-Matthews decimation directly \cite{NHK06}.

%===========================================================================
%===========================================================================

\end{document}